\documentclass[aps,pre,twocolumn,showpacs,preprintnumbers,amsmath,amssymb]{revtex4}

\usepackage{graphicx}
\usepackage{dcolumn}
\usepackage{bm}

\begin{document}


\title{Dissipation, Generalized Free Energy, and a Self-consistent
Nonequilibrium Thermodynamics of Chemically Driven Open Subsystems}

\author{Hao Ge$^{1}$}
\email{haoge@pku.edu.cn}
\author{Hong Qian$^{2}$}%
\email{hqian@u.washington.edu}
\affiliation{$^1$Beijing International Center for Mathematical Research (BICMR) and Biodynamic Optical Imaging Center (BIOPIC), Peking University, Beijing, 100871, PRC.\\
$^2$Department of Applied
Mathematics, University of Washington, Seattle, WA 98195, USA.}

\date{\today}

\begin{abstract}
Nonequilibrium thermodynamics of a system situated in a sustained environment
with influx and efflux is usually treated as a subsystem in a larger, closed
``universe''.  It remains a question what the minimally required description
for the surrounding of such an open driven system  is, so that its nonequilibrium
thermodynamics can be established solely based on the internal stochastic
kinetics.  We provide a solution to this problem using insights from studies
of molecular motors in a chemical nonequilibrium steady state (NESS) with
sustained external drive through a regenerating system, or in a quasi-steady
state (QSS) with an excess amount of ATP, ADP, and Pi.    We introduce the
key notion of  {\em minimal work} that is needed, $W_{min}$, for the external
regenerating system to sustain a NESS ({\em e.g.}, maintaining constant
concentrations of ATP, ADP and Pi for a molecular motor).  Using a Markov (master-equation)
description of a motor protein, we illustrate that the NESS and QSS have
identical kinetics as well as the Second Law in terms of a same positive
entropy production rate.  The difference between the heat dissipation of a
NESS and its corresponding QSS is exactly the $W_{min}$.  This provides a justification for introducing an {\em ideal external regenerating system} and
yields a {\em free energy balance equation} between the net free
energy input $F_{in}$ and total dissipation $F_{dis}$ in an NESS: $F_{in}$ consists of chemical input minus mechanical output; $F_{dis}$ consists
of dissipative heat; and the amount of useful energy becoming heat
is the NESS entropy production.
Furthermore, we show that for non-stationary systems, the $F_{dis}$ and $F_{in}$ correspond to the entropy
production rate and housekeeping heat in stochastic thermodynamics, and identify
a relative entropy $H$ as a generalized free energy.  We reach a new formulation of Markovian nonequilibrium thermodynamics
based on only the internal kinetic equation without  further reference to
the intrinsic degree of freedom within each Markov state.  It includes an extended
free energy balance and a Second Law which are valid for driven stochastic
dynamics with an ideal external regenerating system.  Our result suggests new ingredients for a generalized
thermodynamics of self-organization in driven systems.
\end{abstract}

\pacs{}
\maketitle

\section{Introduction}

Statistical thermodynamics is the mathematical foundation of the material
world in terms of classical physics, on which modern chemistry and biology
is based \cite{thermo_np,thermo_hill,thermo_bq}.  To address the
fundamental issues in complex living organisms such as a cell, there are
currently two rather different perspectives: A classical physicist maintains
a world following the Clausius-Boltzmann's Second Law of
Thermodynamics and considers a living organism as a
{\em subsystem} in a quasi-stationary state (QSS), due to
the slow-changing nature of its environment.  According to this
view, the thermodynamic origin of a living system resides in
the fluctuations of its environment.   On the other hand, engineers and
cellular biologists consider a complex system in a sustained
environment that has to be maintained. How to maintain such
environment is not a concern to someone who is only interested
in the internal, complex kinetics.  In equilibrium
statistical physics, these two perspectives have yielded respectively
Boltzmann's microcanonical and Gibbsian canonical ensemble theories
of matters \cite{textbooks}.

	For isothermal but chemically nonequilibrium systems, this distinction can
be best illustrated by two types of laboratory experiments on a single motor protein \cite{Motor_ki,Motor_oster,Motor_qian,Motor_kf} which converts
chemical energy from ATP hydrolysis in an aqueous solution to
mechanical work at the sub-cellular level. In the first type of experiments,
the amount of ATP, ADP and Pi in the solution
are not controlled. However, due to the excess nature of their amount in
solution, their concentrations can be considered approximately
constant over the entire duration of a single-molecule experiment.
Nevertheless, if an experiment is prolonged for a significant period of
time, the ATP and ADP+Pi will eventually reach their chemical
equilibrium, and the motor protein will cease to execute a
directional motion. In the second type of experiments, an
ATP-regenerating system is coupled to the motor protein
\cite{regen_atp_1,regen_atp_2}.  In this case, the motor protein, as
an open, driven chemical system, can reach a nonequilibrium steady
state (NESS) \cite{JQQ,QH} with cyclic conformational kinetics
\cite{hill_cycle_kinetics} while continuously moves along its track, even
in the presence of a mechanical load.

In the stochastic, kinetic theories of single motor proteins \cite{Motor_oster,Motor_qian,Motor_kf}, both the QSS and NESS
are treated mathematically by assuming time-independent, constant
concentrations of ATP, ADP and Pi, which leads to identical predictions
of the kinetics.  The changes in the ATP, ADP, and
Pi concentrations in QSS are so miniscule, they can be safely neglected.

From the thermodynamic point of view, the Markovian transition rates for a complete conformational cycle of a
single motor protein, say with totally $n$ states, no matter in QSS or NESS, satisfy
(see below and also \cite{efficiency_1})
\begin{eqnarray}
         &&  k_BT\ln\left(\frac{k_{1\rightarrow 2}k_{2\rightarrow 3}
               \cdots k_{n\rightarrow 1}}{k_{2\rightarrow 1}k_{3\rightarrow 2}
               \cdots k_{1\rightarrow n}}\right)
\nonumber\\
           &=& \Delta\mu_{ATP\rightarrow ADP+Pi}-W_{\textrm{mechanical}}
\label{eq001}\\
         &=&\Delta\mu^o_{ATP\rightarrow ADP+Pi}
                       +k_BT\ln\frac{[ATP]}{[ADP][Pi]} - W_{\textrm{mechanical}},
\nonumber
\end{eqnarray}
which is precisely the chemical free energy of a single ATP hydrolysis
minus the amount of motor mechanical energy output
\cite{thermo_hill,QH,Motor_qian,Motor_kf}, i.e. the {\em net amount of
free energy} input over the cycle.  When this net amount of free energy input is
zero, the internal kinetics satisfies detailed balance \cite{hq_decomp}.

	A few remarks to (\ref{eq001}) are in order.
First, we note that for a complete kinetic cycle, the affinity
\begin{subequations}
\begin{eqnarray}
      \gamma &\equiv& \frac{k_{1\rightarrow 2}k_{2\rightarrow 3}\cdots
              k_{n\rightarrow 1}}{k_{2\rightarrow 1}k_{3\rightarrow 2}\cdots
                 k_{1\rightarrow n}}
\\
      &=&  \frac{c_1(t)k_{1\rightarrow 2}}{c_2(t)k_{2\rightarrow 1}}\times
                \frac{c_2(t)k_{2\rightarrow 3}}{c_3(t)k_{3\rightarrow 2}}\times \cdots
                \times \frac{c_n(t)k_{n\rightarrow 1}}{c_1(t)k_{1\rightarrow n}}
\\
     &=&  \frac{c_1^{ss}k_{1\rightarrow 2}}{c_2^{ss}k_{2\rightarrow 1}}\times
                \frac{c_2^{ss}k_{2\rightarrow 3}}{c_3^{ss}k_{3\rightarrow 2}}
                \times \cdots
                \times \frac{c_n^{ss}k_{n\rightarrow 1}}{c_1^{ss}k_{1\rightarrow n}},
\end{eqnarray}
\label{eq002}
\end{subequations}
in which $c_i(t)$ is the concentration of the motor protein
in state $i$ at time $t$, and $c_i^{ss}$ is its steady-state concentration,
assuming the protein solution is ideal, {\em e.g.,} single
molecular kinetics are statistically identical and independent.
In terms of classical chemical thermodynamics
in the absence of a mechanical force ($W_{mechanical}=0$),
the (\ref{eq002}a)  corresponds to the free energy difference of one
ATP hydrolysis, which is independent of the motor protein. Every term in (\ref{eq002}b)
and (\ref{eq002}c) corresponds to the free energy difference of each chemical transformation step including conformational change within this cycle, as the protein is in a transient state or an steady state. But at the level of a complete
cycle with the motor protein returning to its beginning state,  all the concentration terms
drop out and they all become the same. Therefore, as repeated pointed out by
T.L. Hill \cite{thermo_hill,hill_cycle_kinetics,cycle_complete}, the net amount of free
energy dissipation on a cycle level is unambiguous and constitutes the entropy production.

	Second, Eq. (\ref{eq002}) clearly shows that just as each individual
reaction is the fundamental ``unit'' of a complex chemical equilibrium,
each kinetic cycle is the fundamental ``unit'' of a chemical NESS.
This insight has been discussed extensively in \cite{hill_cycle_kinetics,JQQ},
which include a mathematical theorem of cycle decomposition at NESS. In a NESS at the cycle level, the free energy input and free energy dissipation
are also balanced.  The steady state flux distribution among the cycles
provides the probability ``weight'' for the stochastic kinetics.

    Third,  we see that the internal stochastic kinetics does not differentiate
between the amount of chemical energy input and the negative
amount of mechanical output; only the {\em net amount of free energy} input.
This confirms the statement that ``one needs to know more than subsystem
kinetics to deal with the full First Law'' \cite{Seifert2011,Sekimoto}.
In fact, the traditional concept of efficiency, which is observed
by an ``outsider'' to the system, can not be determined from the
internal kinetics alone. It indicates, however, that if an outside
agent can differentiate the chemical input
and the mechanical outout, then the efficiency of the subsystem at NESS
has the appropriate upper-bound:
\begin{equation}
            \frac{\textrm{mechanical output}}{\textrm{chemical input}}
          = 1 - \frac{\textrm{entropy production}}{\textrm{chemical input}}
          \le 1.
\label{eq_003}
\end{equation}
Positive entropy production, thus, is the origin of less-than-100\% efficiency.
All the complications in maintaining the NESS, of course, contribute
to a positive entropy production rate, thus a lower efficiency.
When a subsystem is in equilibrium with its surrounding, the efficiency is 1,
but its actual power is zero. See a discussion of this singular
problem in terms of futile cycles \cite{qian_mie} and the recent
studies on efficiency at maximum power \cite{efficiency_2,efficiency_3}.

	A complete discussion of energetics requires further detailed
thermodynamics beyond the level of free energy, which is
decomposed into entropy and enthalpy (or intrinsic energy
 for system with constant volume).  Onto this level, the two setups
NESS and QSS become very different \cite{Seifert2011}: In the QSS, the heat associated with each kinetic cycle is the enthalpy change $\Delta h$
of ATP hydrolysis.   For the NESS, the heat analysis for the ATP-regenerating system is much more complicated.

	The objective of the present paper is two-fold:  First, in Sec.
\ref{sec2}, a more concrete and detailed energetic comparison, including the First Law and
reaction heat, is carried out for the QSS and NESS. With an
additional piece of information concerning what exactly the regenerating system does, we show that
Eq. (\ref{eq001}) is also the {\em minimal amount} of heat dissipated possible to sustain the NESS.  We shall call a regenerating system with the minimal
heat dissipation {\em ideal}. Equipped with this novel notion, we show that
the entropy production of a Markov process defines the
amount of dissipated ``heat'', and the equation in (\ref{eq001})
is a form of ``free energy balance'' at NESS, e.g., analogous to the First Law.
Our theory shows consistency and contradistinctions in thermodynamics of energy transduction and heat dissipation in the two different perspectives of nonequilibrium systems, QSS and NESS {\em \`{a} la} Clausius and Kelvin.

Then in Sec. \ref{sec3}, for non-stationary
process, we can further generalize the free energy balance equation
and identify a generalized free energy, together with a Second Law. When there is only internal stochastic dynamical
information, without any detailed knowledge about the intrinsic degrees of freedom of each Markov discrete state, as in many applications of Markov models to non-molecular systems,  we have to forgo the traditional
First Law on energy conservation. Interestingly, if we only focus on the level of free energy, we will
get a {\em conservation
law of a generalized free energy}
\begin{equation}
   H\left(\{c_i\}\|\{c_i^{ss}\}\right)
    =k_BT\sum_i c_i\ln \left(c_i/c^{ss}_i\right),
\label{eq_017}
\end{equation}
for a system approaching NESS, which is suggested recently from the mathematical point of view \cite{hq_decomp,Sc,ep_fd_1,ep_fd_2}:
\begin{subequations}
\begin{equation}
     \frac{d}{dt}
     H\left(\{c_i(t)\}\|\{c_i^{ss}\}\right)
    = F_{in}(t) - F_{dis}(t);
\label{eq_018}
\end{equation}
in which
\begin{equation}
      F_{in}(t)=\frac{k_BT}{2}\sum_{ij}
\left(c_ik_{ij}-c_jk_{ji}\right) \ln\left(\frac{c^{ss}_ik_{ij}}
            {c_j^{ss}k_{ji}}\right) \ge 0
\label{eq_19},
\end{equation}
\begin{equation}
    F_{dis}(t) = \frac{k_BT}{2}\sum_{ij}
\left(c_ik_{ij}-c_jk_{ji}\right) \ln\left(\frac{c_i(t)k_{ij}}
            {c_j(t)k_{ji}}\right) \ge 0,
\label{eq20}
\end{equation}
and $F_{dis}(t)$ can be further decomposed into
\begin{equation}
       F_{dis}(t) = T\frac{dS}{dt} + \frac{k_BT}{2}\sum_{ij} \left(c_ik_{ij}-c_jk_{ji}\right) \ln
                       \left(\frac{k_{ij}}{k_{ji}}\right).\label{eq5d}
\end{equation}
\label{eq3}
\end{subequations}
where $S(t)=-k_B\sum_i c_i\ln c_i$ and $k_{ij}$ the transition rate from Markov state $i$ to $j$.
However, the novel mathematical results (\ref{eq3}), only dependent on the internal kinetics,
still beg for a clear thermodynamics interpretations, at least
in the simple example of motor protein.

We will show in Sec. III that the nonnegative $F_{in}(t)$, also called housekeeping heat \cite{ep_fd_1,ep_fd_2,OP,HS}, can be regarded as
the free energy input to sustain the corresponding NESS, while the entropy production rate $e_p(t)=\frac{F_{dis}(t)}{T}$ is just the dissipation at the level of free energy. They are equal to each other in steady state. These two important quantities are all independent of intrinsic degrees of freedom of each chemical/conformational state.

Hence Eq. (\ref{eq_018})
is most reasonably interpreted as:

\begin{center}
\bf\em
change in the generalized free energy  \boldmath{$=$} source

(free energy and work) \boldmath{$-$} sink (dissipation).
\end{center}
\noindent
This is a law of balance for the extensive quantity $H$.
One can understand the $H$ as a generalized free energy.

The function $H$ indicates how far the system deviates from the sustained NESS \cite{qian_pre_01}. When a system is very far from
its NESS, it has a large $H$, then the non-stationary condition
of the system constitutes a source of entropic force
which can be utilized; and $H=0$ when a system is at its NESS.

	Based on this newfound NESS perspective, in Sec. \ref{sec3}
part we shall also show that the new perspective further yields an extended Second Law, which emerges only from driven kinetics with an idealized external regenerating system.

\section{Energetic comparison and nonequilibrium thermodynamics between QSS and NESS}
\label{sec2}

The nonequilibrium thermodynamical analysis of the QSS is quite traditional.
It follows the original idea of Boltzmann and it is generally applicable: Heat, work, entropy, free energy and other thermodynamic quantities are all well defined. Unfortunately, such an analysis could not be carried out solely based on the internal kinetics of the subsystem; one needs to know nearly every detail of the surroundings {e.g. particles in the solution and their
 interactions} \cite{Seifert2011}.   Furthermore, except with extensive
studies on temperature dependence, what is heat is never unambiguous
\cite{benzinger}.  Therefore for the engineers or cellular biologists who are only interested in the internal kinetics of the motor protein, a self-consistent
and nearly self-contained thermodynamics is highly desirable.  In this part
we develop the NESS perspective and try to deduce important thermodynamic relations from internal stochastic kinetics, plus
additional information:  ($A$) The identification of the input free energy and output work.  Usually for motor the input is chemical free
energy and the output is mechanical work. ($B$) Further decomposition
of free energies into their entropic and enthalpic parts.

\begin{figure}[h]
\includegraphics[width=3.cm,angle=270]{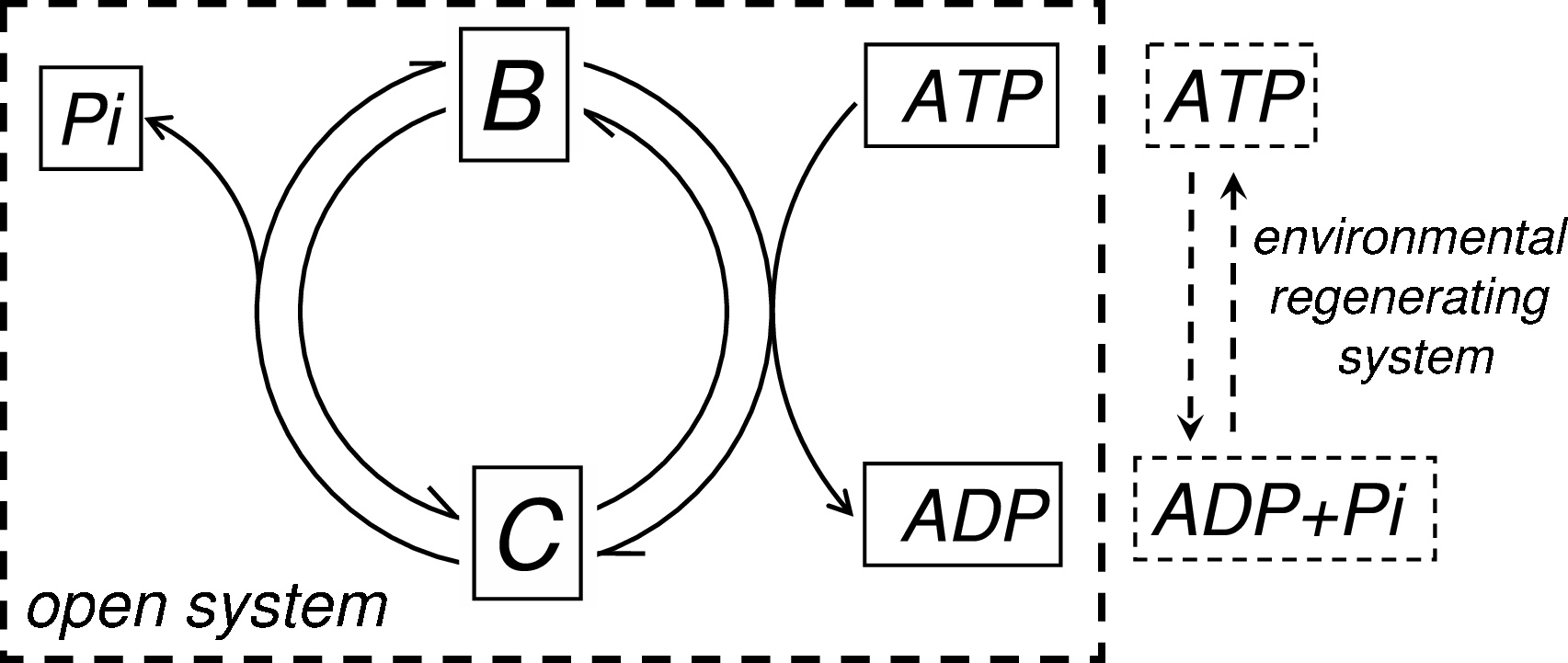}
\caption{The thermodynamics of spontaneous ATP hydrolysis and
related ATP regenerating process.  The entire cycle would have
an $\Delta\mu>0$ amount of net free energy input and the same amount of energy dissipated.  $\Delta\mu$ is just the free energy change of the ATP hydrolysis minus the amount of mechanical energy output.  See the main text for details.} \label{fig1}
\end{figure}

\subsection{Minimal external work, ideal regenerating system, and
an energy balance equation for a subsystem}

Fig. \ref{fig1} shows a simple biochemical reaction
cycle $B\rightarrow C\rightarrow B$ coupled to ATP hydrolysis.
The ATP, ADP and Pi concentrations are maintained by an
``external'' regenerating system:
\begin{equation}
     B+ATP\overset{k_1}{\underset{k_{-1}}{\rightleftharpoons}}
           C+ADP,\ \
   C\overset{k_2}{\underset{k_{-2}}{\rightleftharpoons}} B+Pi.
\label{basic_rxn}
\end{equation}
After completing a reaction cycle (\ref{basic_rxn}),
the net effect is one ATP being hydrolyzed to ADP+Pi.
At the meantime, the regenerating system would
convert ADP+Pi back to ATP externally. This is the
essential difference between NESS and QSS which results
in one ATP hydrolysis after one cycle.
Standard thermodynamics tells us the chemical potentials
of each species are defined as
\begin{eqnarray}
&& \mu_B=\mu_B^o+k_BT\ln [B],
   ~~\mu_C=\mu_C^o+k_BT\ln [C],
\nonumber\\
&& \mu_{ATP}=\mu_{ATP}^o+k_BT\ln [ATP],
\nonumber\\
&&  \mu_{ADP}=\mu_{ADP}^o+k_BT\ln [ADP],
\nonumber\\
&&   \mu_{Pi}=\mu_{Pi}^o+k_BT\ln[Pi].
\label{eqn_007}
\end{eqnarray}
At chemical equilibrium, $\mu_B+\mu_{ATP}=\mu_C+\mu_{ADP}$ and
$\mu_B+\mu_{Pi}=\mu_C$, i.e.
$k_1[B]^{eq}[ATP]^{eq}=k_{-1}[C]^{eq}[ADP]^{eq}$ and
$k_2[C]^{eq}=k_{-2}[B]^{eq}[Pi]^{eq}$, which also leads to the
thermodynamic relations
\begin{eqnarray}
   \mu_B^o+\mu_{ATP}^o-\mu_C^o-\mu_{ADP}^o &=& k_BT\ln
      \left(k_1/k_{-1}\right),
\label{eqn_008}\\
  \mu_C^o-\mu_B^o-\mu_{Pi}^o &=& k_BT\ln\left(k_2/k_{-2}\right).
\label{eqn_009}
\end{eqnarray}
For a complete cycle, one combines  Eqs. (\ref{eqn_007}),
(\ref{eqn_008}) and (\ref{eqn_009}).  The $\mu_B$'s and $\mu_C$'s
in the two reactions in (\ref{basic_rxn}) cancel out, and one is left with
\begin{eqnarray}
	 \Delta\mu_{ATP\rightarrow ADP+Pi}   &\equiv&
			  \mu_{ATP}-\mu_{ADP}-\mu_{Pi}
\nonumber\\
	 &=& k_BT\ln\frac{k_1k_2[ATP]}{k_{-1}k_{-2}[ADP][Pi]},
\label{eqn_010}
\end{eqnarray}
a special case of the Eq. (\ref{eq001}).

Each intrinsic chemical potential can be further decomposed into
$\mu^o=h^o-Ts^o$, where $h^o$ and $s^o$ are the intrinsic enthalpy
and entropy respectively. Then for a single occurrence of the
hydrolysis cycle in Fig. \ref{fig1}, the heat dissipation
is
\begin{eqnarray}
Q_d &=& (h_B^o+h_{ATP}^o-h_C^o-h_{ADP}^o)+(h_C^o-h_{Pi}^o-h_B^o)
\nonumber\\
  &=&  h_{ATP}^o-h_{ADP}^o-h_{Pi}^o.
\end{eqnarray}
Note that the $Q_d$ can be negative, or even greater than the
hydrolysis free energy in (\ref{eqn_010}).  This portion of the
energy is stored in chemical bond of an ATP molecule.

With the presence of a regenerating system, there is an ``external
step'' converting ADP+Pi back to ATP after each completion of an
enzyme cycle.  The minimum work (non-PV work) it has to do is the
free energy difference between ADP+Pi and ATP, i.e.
\begin{equation}
             W_{min}=\mu_{ATP}-\mu_{ADP}-\mu_{Pi}
                 \equiv \Delta\mu_{ATP\rightarrow ADP+Pi},
\end{equation}
with corresponding enthalpy changes $-Q_d$.
Therefore, the energy dissipation of this external step in the environment, in the form of heat, is
\begin{equation}
  Q_d^{ext}=W_{min}-(h^o_{ATP}-h^o_{ADP}-h^o_{Pi}).
\end{equation}
We note that $Q^{ext}_d$ is just the entropy change, off
by a temperature factor, for the ATP hydrolysis.

Hence the {\em total} heat dissipation of a single forward biochemical
cycle in a driven system with regeneration is
\begin{eqnarray}
Q_{tot}&=&Q_d+Q_d^{ext}\nonumber = W_{min}
     =\Delta\mu_{ATP\rightarrow ADP+Pi}\\
  &=& k_BT\ln\gamma,
\end{eqnarray}
where $\gamma=\frac{k_1k_2[ATP]}{k_{-1}k_{-2}[ADP][Pi]}>1$ is
the affinity for the reaction cycle.

	If, however, the regenerating system is not perfectly efficient, then
$W_{actual}>W_{min}$ and part of it is wasted in the process:
$\delta=W_{actual}-W_{min}$.  It follows that
$Q_d^{ext}=$ $W_{actual}-Q_d=$ $W_{min}-Q_d+\delta=$
$k_BT\ln\gamma + \delta$.   We call a regenerating system with $\delta=0$
ideal.  Therefore, with the assumption of an ideal regenerating system, a balance
between chemical free energy and heat can be estabished,
on the level of kinetic cycles.

Now if there is an external, mechanical force exerted
on the motor, then the minimum external work to sustain the NESS is still $W_{min}=\Delta\mu_{ATP\rightarrow ADP+Pi}$ \cite{footnote_001}, but $\Delta\mu_{ATP\rightarrow ADP+Pi}$ is no longer equal to
$k_BT\ln\gamma$. Their difference is the mechanical
output of the system, i.e. $Fd$, where $F$ is the
force and $d$ is the step size of the motor with one
ATP hydrolysis. Hence  Eq. (\ref{eq001}) becomes a form of
energy balance.
When $\gamma>1$, all the chemical input minus the mechanical output
is dissipated while the system remaining steady.

We see the central importance of cycle kinetics from this simple
example. Before a completion of a cycle, the regenerating system
needs not to do anything to maintain the environment, and all the
work done to ``the system'' is potentially reversible.  This has
been emphasized by T.L. Hill \cite{cycle_complete}; a similar
argument was put forward by R. Landauer for the thermodynamics of computation \cite{landauer}.

\subsection{Master equation system and thermodynamic constrains}

The above results for a single biochemical cycle can be generalized to
dynamical models with master equations: Let us consider a motor
protein with $N$ different conformations $R_1,R_2,\cdots,R_N$.
Suppose that the system is kept in a close contact with a large heat
bath with constant temperature $T$ and pressure.
For simplicity, the concentration of every substance is assumed to
be independent of its position, and there is no external input or
output of mechanical energy.  Introducing the mechanical
part is straightforward as illustrated for the single cycle above.

Let $k_{ij}$ be the first-order, or pseudo-first-order
rate constants for reaction $R_i\rightarrow R_j$.  Assume only one of
them is coupled with a chemical free energy source, i.e., ATP and ADP:
$$ATP+R_1\overset{\widetilde{k}_{12}}{\underset{\widetilde{k}_{21}}{\rightleftharpoons}}ADP+R_2,$$
where $\widetilde{k}_{12}$ and $\widetilde{k}_{21}$ are both second-order
reaction constants, and $k_{12}=\widetilde{k}_{12}[ATP]$,
$k_{21}=\widetilde{k}_{21}[ADP]$ are pseudo-first-order rate
constants. For simplicity, here we omit the Pi release step, since what
we need here is only the reaction $R_1\rightarrow R_2$  having a driving force.

Let $c_i$ be the concentration of $R_i$. Then by the law of mass
action, such a linear system could be described in terms of a
mathematical model
\begin{equation}\frac{dc_i(t)}{dt}=\sum_{j}
\left(c_j k_{ji} -c_ik_{ij}\right). \label{master_eq}
\end{equation}

If there is no external mechanism to keep the concentrations of
ATP and ADP, then the time evolution of $c_T=$[ATP] and
$c_D=$[ADP] is
\begin{equation}
\frac{dc_T}{dt}=-\frac{dc_D}{dt}=-\widetilde{k}_{12}c_Tc_1+\widetilde{k}_{21}c_Dc_2.\label{TD_eq}
\end{equation}
Classical equilibrium thermodynamics for closed chemical system
tells us that there is a unique dynamic and chemical equilibrium
$\{c_1^{eq},c_2^{eq},\cdots,c_N^{eq},c_T^{eq},c_D^{eq}\}$ which
satisfies the detailed balance condition
$c_i^{eq}k_{ij}=c_j^{eq}k_{ji}$,
where $k_{12}=\widetilde{k}_{12}c_T^{eq}$ and
$k_{21}=\widetilde{k}_{21}c_D^{eq}$.

Each species has a chemical potential $\mu_i(c_i) = \mu_i^o+k_BT\ln
c_i$, where $\mu_i^o$ is the internal chemical potential
of species $R_i$ and obeys the Boltzmann's law $\mu_i^o= -k_BT\ln
c_i^{eq}+$const. When a system reaches chemical equilibrium, the chemical
potentials of different components are the same, i.e.
$\mu_i(c^{eq}_i)=\mu_j(c^{eq}_j)$, and
$\mu_1(c^{eq}_1)+\mu_T(c^{eq}_T)=\mu_2(c^{eq}_2)+\mu_D(c^{eq}_D),$
where $\mu_T(c_T)=\mu_T^o+k_BT\ln c_T$ and
$\mu_D(c_D)=\mu_D^o+k_BT\ln c_D$ are the chemical potentials of ATP
and ADP respectively.

Then it gives the relation between $\mu^o$'s and
$k_{ij}$'s of the system, i.e.
\begin{eqnarray}
&& \mu_i^o-\mu_j^o=k_BT\ln\frac{k_{ij}}{k_{ji}}, \ \ \mu_T^o-\mu_D^o
=k_BT\ln\frac{c_D^{eq}}{c_T^{eq}},
\nonumber\\
&&\mu_1^o+\mu_T^o-\mu_2^o-\mu_D^o=
k_BT\ln\frac{\widetilde{k}_{12}}{\widetilde{k}_{21}}.\label{detailedbalance}
\end{eqnarray}

\subsection{Thermodynamics of subsystem QSS within a larger closed system
with detailed balance}

In this case, the whole system is closed including ATP and ADP;
its final dynamical equilibrium is a chemical equilibrium.
The total free energy of the system with concentrations $c_i$,
$c_D$, and $c_T$, is
\begin{equation}
F^{close}=\sum_ic_i\mu_i+c_T\mu_T+c_D\mu_D.\nonumber
\end{equation}
$F^{close}(t)$  always decreases until it reaches to its minimum at
equilibrium:
\begin{equation}
\frac{dF^{close}(t)}{dt}
=-k_BT\sum_{i>j}(c_ik_{ij}-c_jk_{ji})\ln\left(
          \frac{c_ik_{ij}}{c_jk_{ji}}\right)\leq
0. \label{eq_15}
\end{equation}
The term
$f_d^{close}=-dF^{close}(t)/dt$ is called {\em free energy
dissipation rate} \cite{ep_fd_1}.

For each chemical/conformational state $i$ of the motor protein, the internal entropy $Ts^o_i=h_i^o-\mu_i^o$.
Thus the entropy of the entire system could be defined as
$\widetilde{S}^{close}=S^o+S^{close}$, where $S^o=\sum_i s_i^oc_i+s_T^oc_T+s_D^oc_D$ and
$S^{close}=k_B\sum_i[-c_i\ln c_i]-c_T\ln c_T-c_D\ln c_D$. Then the evolution of entropy
becomes
\begin{equation}
   \frac{d\widetilde{S}^{close}}{dt}=e_p^{close}-\frac{\widetilde{h}_d^{close}}{T},
\label{eq_012}
\end{equation}
where
\begin{eqnarray}
\widetilde{h}_d^{close}&=&\frac{1}{2}\sum_{ij}(c_ik_{ij}-c_jk_{ji})(h_i^o-h_j^o)\nonumber\\
&&+(c_1k_{12}-c_2k_{21})(h_T^o-h_D^o)\nonumber
\end{eqnarray}
is the heat dissipation, and the {\em entropy production rate}
$Te_p^{close}=f_d^{close}$  \cite{thermo_np,thermo_hill,Sc,JQQ}.
The entropy of the system increases due to entropy
generated in spontaneous processes and decreases when
heat is expelled into the surrounding.  Eq. (\ref{eq_012}) is an example
of the entropy balance equation of Dutch School's nonequilibrium
thermodynamics $dS/dt=d_iS/dt+d_eS/dt$ \cite{degroot,thermo_np}.

If one only regards the motor protein as the unique target system, and define $\widetilde{S}^{motor}=\sum_i s_i^oc_i-k_B\sum_ic_i\ln c_i=\widetilde{S}^{close}-\widetilde{S}^{ATP,ADP}$, then we have
\begin{equation}
    \frac{d\widetilde{S}^{motor}}{dt}=e_p^{motor}-\frac{\widetilde{h}_d^{motor}}{T},
\label{dSmotor_dt}
\end{equation}
where $e_p^{motor}=e_p^{close}$, and $\widetilde{h}_d^{motor}=\widetilde{h}_d^{close}+T\frac{d\widetilde{S}^{ATP,ADP}}{dt}$. Here we notice that the definition of entropy and entropy production rate are independent of the mechanical details of the environment, but the traditional definition of heat, i.e., $\widetilde{h}_d^{close}$, is.  It involves the entropy change in the solution resulting from the reaction \cite{Seifert2011}. One needs to overcome such a non-self-containment in order to reach a thermodynamic framework solely based on the internal kinetics of the target system, which is at NESS.

\subsection{Thermodynamics of nonequilibrium driven system}

With the presence of an {\em ideal} external regenerating mechanism,
the concentrations of ATP and ADP would be kept invariant. The system
is not at equilibrium in general \cite{footnote_1}. The internal chemical
kinetics is again described by (\ref{master_eq}), which approaches to
a NESS.
Once the concentrations of $ATP$ and $ADP$ are sustained, since the stochastic kinetics of all the motor proteins are statistically identical
and independent, we could substitute the concentration with probability in (\ref{master_eq}), and talk about stochastic thermodynamics with a
single-molecule perspective \cite{Seifert2011}.

Recall that each $\mu^o$
could be decomposed into $h^o-Ts^o$, hence for each individual
occurrence of the transition $R_i\rightarrow R_j$, the heat
dissipation is $h_i^o-h_j^o$ which is not coupled with the
regenerating system. However, for the real driven reaction $ATP+R_1
\rightleftharpoons ADP+R_2$, the total heat dissipation should be
$(h_1^o+\mu_T)-(h_2^o+\mu_D)$ following the above analysis of the simple example in Fig. \ref{fig1}.

Therefore the heat dissipation rate in such a driven open system
is
\begin{eqnarray}
\nonumber
   \widetilde{h}_d^{open}(t)
       &=&\sum_{i>j} \left(c_i(t)k_{ij}-c_j(t)k_{ji}\right)(h_i^o-h_j^o)\nonumber\\
       &&+\left(c_1(t)k_{12}-c_2(t)k_{21}\right)(\mu_T-\mu_D)\nonumber
\end{eqnarray}
Furthermore, the heat dissipation in the stationary NESS becomes
\begin{eqnarray}
\widetilde{h}_d^{ness}\nonumber
   &=&\sum_{i>j} \left(c_i^{ss}k_{ij}-c_j^{ss}k_{ji}\right)
       (\mu_i^o-\mu_j^o)\nonumber\\
       &&+\left(c_1^{ss}k_{12}-c_2^{ss}k_{21}\right)(\mu_T-\mu_D)\nonumber\\
&=&k_BT\sum_{i>j} \left(c_i^{ss}k_{ij}-c_j^{ss}k_{ji}\right)
       \ln\frac{k_{ij}}{k_{ji}}.
\label{eq_0011}
\end{eqnarray}
The rigorous derivation of (\ref{eq_0011}) is based on the fact that
in an NESS, its kinetics and thermodynamics can be decomposed into
different cycles \cite{thermo_np,thermo_hill,Sc,JQQ}. As we have stated, the
regenerating system would not really do any irreversible work unless
there is a completion of a driven cycle. The amount of minimum work,
done by the ideal regenerating system, for each internal
cycle $c=\{i_0\rightarrow
i_1\rightarrow i_2\cdots\rightarrow i_n\rightarrow i_0\}$ is
$$W_{min}^c=k_BT\ln\frac{k_{i_0i_1}k_{i_1i_2}\cdots k_{i_ni_0}}{k_{i_0i_n}k_{i_ni_{n-1}}\cdots k_{i_1i_0}},$$
which is also equal to the total heat dissipation $Q_{tot}^c$ for
the same cycle.

For each state $i$, the internal entropy $Ts^o_i=h_i^o-\mu_i^o$.
Thus the entropy of the open system could be defined as
$\widetilde{S}^{open}=S^o+S^{open}$, where $S^o=\sum_i s_i^oc_i$ and
$S^{open}=-k_B\sum_i c_i\ln c_i$. The evolution of entropy, thus,
\begin{equation}
   \frac{d\widetilde{S}^{open}}{dt}=e_p^{open}-\frac{\widetilde{h}_d^{open}}{T},
\end{equation}
where $e_p^{open}=k_B\sum_{i>j} \left(c_ik_{ij}-c_jk_{ji}\right)
       \ln\frac{c_ik_{ij}}{c_jk_{ji}}$ is the entropy production
rate \cite{thermo_np,thermo_hill,Sc,JQQ}. The ``heat term'' $\widetilde{h}_d^{motor}(=\widetilde{h}_d^{open})$ in Eq. (\ref{dSmotor_dt}) now finally becomes real heat and is completely
independent of the any details on the regenerating system.

One could easily notice that the entropy $\widetilde{S}^{open}=\widetilde{S}^{motor}$, and more important
$Te_p^{open}=Te_p^{motor}=Te_p^{close}=f_d^{close}$. It indicates that the entropy production rate is indeed independent of the QSS or NESS
perspectives of the subsystem. More importantly, this shows a
consistency between the different
perspectives of Boltzmann/Gibbs and Prigogine for the traditional
Second Law: Boltzmann states entropy never decreases in an
isolated system and Gibbs states free energy never increases in
a closed isothermal system; while Prigogine states that the entropy
production is never negative in an open system, and it can be defined
solely from internal kinetics. They are equivalent.

Thus the free energy of the open system
$$\widetilde{F}^{open}=h^o-T\widetilde{S}^{open}=\mu^o-TS^{open},$$
where $h^o=\sum_i h^o_ic_i$ is the enthalpy, and
$\mu^o=\sum_i\mu_i^oc_i$ is the internal (conditional)
free energy of the system. Note that no matter how large the entropic component of $\mu^o_i$
is, $Ts_i^o$ enters both $h^o$ and $T\widetilde{S}^{open}$ and they
compensate, leaving $\widetilde{F}^{open}$ invariant
\cite{santillan_qian}. The evolution of such a free energy function would not always decrease any more, which spurred the discovery of relative entropy as a generalized free energy for NESS (see below).

\subsection{Housekeeping heat: the driver of NESS}

In the NESS perspective, the heat dissipation for the transition $i\rightarrow j$ is
$Q_{ij}=k_BT\ln\left(k_{ij}/k_{ji}\right)+T(s_i^o-s_j^o)$
no matter coupled with the driving force or not; and meanwhile, the steady-state entropy also could be defined for this single transition \cite{Seifert05} as $\Delta S_{ij}^{ss}=k_B\ln\left(c_{i}^{ness}/c_{j}^{ness}\right)+(s_j^o-s_i^o)$. Therefore, the housekeeping heat $Q_{hk}=F_{in}$ in (\ref{eq_19}) is just ensemble averaged difference between $T\Delta S_{ij}^{ss}$ and $Q_{ij}$, which is equal to the entropy production rate at NESS.

Housekeeping heat $Q_{hk}$ is really the nonequilibrium driver of the system. $Q_{hk}(t)\equiv 0$ if and only if $T\Delta S_{ij}^{ss}+Q_{ij}=0$, which matches the classic definition of equilibrium entropy difference through a reversible process. Hence for master equation with detailed balance which correspond to
closed system, $Q_{hk}(t)\equiv 0$ whenever the system is in the steady or any transient state.

Some might argue that the definition of steady-state entropy contains ensemble information, hence it could not be defined along a stochastic path. We think it is indispensable to distinguish the stationary and instantaneous concentration/distribution of the system. Although the former one still could be regarded as an ensemble property, it is {\em measurable} through {\em ergodic} internal kinetics, which is only dependent on molecular structures as well as solvent concentrations. Hence, it could somehow still be considered as ``intrinsic'' property,
{\em without a need for an ensemble picture}.

\subsection{Efficiency for the chemical to mechanical
energy transduction in a NESS}

A molecular motor is a mechanical system coupled fully reversibly to
a chemical reaction or reactions, with an external force $F_{mechical}$
resisting the mechanical movement driven by the chemical gradient.
The external force $F_{mechanical}$ does not effect the
$\Delta\mu_{ATP\rightarrow ADP+Pi}$, which is still equal to $W_{min}$. Rather, the logarithmic affinity in Eq. (\ref{eq001}) now contains a
chemical part and a mechanical part.   The $F_{mechanical}$,
therefore, reduces the amount of entropy production as well as
the dissipation.  However, when the $F_{mechanical}$ is greater
than the stalling force, the entropy production again increases
and the mechanical energy is now being converted into chemical
potential, i.e., the chemical flux is against the chemical
potential in ATP synethesis.  This scenario has been realized in the reversely run
F$_0$F$_1$ ATPase becoming a ATP synthetase \cite{Motor_ki,Motor_oster}.
On the other hand, the reversal of chemomechanical energy transduction
would not occur if the coupling is not fully reversible.  A
load from a viscous drag force is an example.  The direction of the force is always
against the stochastic flux that generates movements.

Quantitatively, in the case of chemical to mechanical
transduction, the energy conservation is
$W_{min}\times J_{c\rightarrow m}=T\cdot e_p^{open}+P_{mech}$
where $P_{mech}$ is a mechanical power, and the
efficiency $\eta=\frac{P_{mech}}{T\cdot e_p^{open}+P_{mech}}\leq 1$.
In the opposite direction, the above $P_{mech}$ and
$J_{c\rightarrow m} <0$.
Hence, $\eta=\frac{W_{min}\times |J_{c\rightarrow
m}|}{W_{min}\times |J_{c\rightarrow m}|+T\cdot e_p^{open}}\leq 1$.
We see that whether the energy transduction is
chemical to mechanical or the opposite, the entropy
production is always the total dissipation and it is
nonnegative, resulting in a less than $100\%$ efficiency. Although such an expression of energy transduction efficiency has been used in many previous works \cite{efficiency_1,efficiency_2}, its physical meaning becomes more clear now.

\section{Self-consistent nonequilibrium thermodynamics at the level of free energy}

\label{sec3}

In reality, we usually know little about internal degrees of freedom of each
Markov state, especially as we study the biological processes. Enlightened by the detailed comparison of QSS and NESS systems as well as the thermodynamic relation (\ref{detailedbalance}), we realized that the internal kinetics of the system, e.g. the transition between different conformational states of the motor protein, is essentially related to free energy rather than the intrinsic enthalpy or entropy of each chemical/conformational state.
In statistical chemistry, this is the notion of ``conditional free energy'' for
discrete states, or ``potential of mean force'' for continuous varables
\cite{santillan_qian}.

We now propose a self-consistent nonequilibrium thermodynamics of a chemically driven open system with sustained surroundings at the level of free energy, only based on the internal kinetics of the system, which is in a time-dependent transient state toward the corresponding NESS.

\subsection{Intrinsic free energy dissipation}

At the level of free energy, one can first define the {\em intrinsic free energy dissipation}
\begin{eqnarray}
\nonumber
   F_{dis}^o(t)
       &=&\sum_{i>j} \left(c_i(t)k_{ij}-c_j(t)k_{ji}\right)\Delta \mu^o_{ij}\nonumber\\
       &&+\left(c_1(t)k_{12}-c_2(t)k_{21}\right)(\mu_T-\mu_D)\nonumber\\
       &=&k_BT\sum_{i>j} \left(c_ik_{ij}-c_jk_{ji}\right)
       \ln\frac{k_{ij}}{k_{ji}},
\end{eqnarray}
where $\Delta \mu^o_{ij}=\mu_i^o-\mu_j^o$ is the intrinsic free energy difference between the states $i$ and $j$.
At steady state, it is just equal to the heat dissipation $\widetilde{h}_d^{open}$, which implies the thermodynamics at the level of free energy or at more detailed level of intrinsic enthalpy and entropy are the same at NESS. However, the heat dissipation at the transient state is dependent on the decomposition of free energy of each chemical/conformational state into entropy and enthalpy, which is beyond the level of free energy. Hence now we shall use the intrinsic free energy dissipation instead of the heat dissipation.

\subsection{Entropy production as the free energy dissipation}

In the case of QSS, we already know that $Te_p^{close}=f_d^{close}$, which implies the entropy production rate is ultimately related to the free energy dissipation. Still, in the open driven system, we can also interpret the entropy production rate $Te_p^{open}$ as
\begin{eqnarray}
\nonumber
   Te_p^{open}(t)
       &=&\sum_{i>j} \left(c_i(t)k_{ij}-c_j(t)k_{ji}\right)\Delta \mu_{ij}\nonumber\\
       &&+\left(c_1(t)k_{12}-c_2(t)k_{21}\right)(\mu_T-\mu_D)\nonumber\\
       &&=k_BT\sum_{i>j} \left(c_i k_{ij}-c_j k_{ji}\right)
       \ln\frac{c_i k_{ij}}{c_j k_{ji}},
\end{eqnarray}
where $\Delta \mu_{ij}=\mu_i-\mu_j$ is the free energy difference between the states $i$ and $j$. Hence $F_{dis}(t)=Te_p^{open}(t)$ is the total free energy dissipation.

\subsection{Housekeeping heat as the free energy input by the external regenerating system}

In the previous sections, we have shown that the housekeeping heat indicates the active driver of a system, e.g., whether an external regenerating system
is present.  If the corresponding steady-state is an equilibrium, i.e. a system
is not externally driven, then the housekeeping heat vanishes for
all time even when the system is in a time-dependent  transient state.

We also notice that,  except $\Delta \mu^{ss}_{12}=\mu_1^{ss}-\mu_2^{ss}+\mu_T-\mu_D$, all other
$T\Delta S_{ij}^{ss}-Q_{ij}=\Delta \mu^{ss}_{ij}$, where $\Delta \mu^{ss}_{ij}=\mu_i^{ss}-\mu_j^{ss}$ is actually the NESS free energy difference along the transition from the state $i$ to $j$, which should be always sustained by the external regenerating system. Therefore, the housekeeping heat $Q_{hk}(t)$ could be regarded as the total free energy input $F_{in}(t)$ to drive such a nonequilibrium system.

In fact, if one knows the entry points for the external free energy input,
one can rewrite $F_{in}$ as
$$F_{in}=-\frac{d\left[\sum_i \mu_i^{ss}p_i\right]}{dt}+(\mu_T-\mu_D)\big(p_1(t)k_{12}-p_2(t)k_{21}\big).$$

\subsection{Generalized free energy and its time evolution}

For a system approaching equilibrium, the free energy input $F_{in}=Q_{hk}$ is zero while the free energy dissipation $Te_p^{close}$ is exactly the derivative of the function $H$ in (\ref{eq_017}), which is actually the free energy deviation from the equilibrium \cite{qian_pre_01},
and then Eq. (\ref{eq_018}) is reduced to (\ref{eq_15}).

For a system approaching NESS, both the free energy input and dissipation are positive. Interestingly, the net free energy dissipation $F_{dis}^{net}(t)=F_{dis}(t)-F_{in}(t)$ is still the derivative of the function $H$ in (\ref{eq_017}). Hence at the level of free energy, we can regard $H$ as a generalized free energy at this specific level, whose time evolution is characterized by (\ref{eq_018}), i.e.
\begin{equation}
     \frac{d}{dt}H\left(\{c_i(t)\}\|\{c_i^{ss}\}\right)=F_{in}(t)-F_{dis}(t).
\end{equation}

\subsection{An extended Second Law and evolution of entropy}

Not only $F_{in}$ and $F_{dis}$ are both nonnegative, $F_{dis}^{net}(t)=-\frac{dH}{dt}=F_{dis}-F_{in}$ is also nonnegative, which vanishes if and only if the system is at steady-state. Hence $F_{dis}^{net}(t)\geq 0$ could be regarded as an extended Second Law, while the traditional Second Law is just $e_p(t)=\frac{F_{dis}(t)}{T}\geq 0$. They are equivalent only when $F_{in}=Q_{hk}=0$, which implies the absence of external regenerating system.

The extended Second Law could also be expressed through the evolution of entropy. Define the entropy of the system as $S(t)=-k_B\sum_i c_i(t)\ln c_i(t)$, we can get Eq. (\ref{eq5d}), i.e.
\begin{equation}
     \frac{dS}{dt}=\frac{F_{dis}-F_{dis}^o}{T},
\end{equation}
which is one specific form of the fundamental entropy balance equation of
nonequilibrium thermodynamics \cite{thermo_np,ep_fd_1,ep_fd_2,Seifert05}. It indicates that the change of entropy is equal to the {\em non-intrinsic free energy dissipation}.

Since the nonnegativity of $e_p(t)=\frac{F_{dis}(t)}{T}$, we then have
$$\frac{dS}{dt}\geq -\frac{F_{dis}^o(t)}{T}.$$
We refer it as the Clausius inequality in the traditional Second Law of thermodynamics, since $F_{dis}^o(t)$ is the same as the heat dissipation, when the system is at NESS.

Furthermore, due to the fact that $F_{dis}(t)=Te_p(t)=F_{in}(t)+F_{dis}^{net}(t)$, we can get
\begin{equation}
     \frac{dS}{dt}=\frac{F_{in}+F_{dis}^{net}-F_{dis}^o}{T}(t).
\end{equation}
Since $F_{dis}^{net}(t)$ is nonnegative, an extended Second Law emerges as
\begin{equation}
     \frac{dS}{dt}\geq \frac{F_{in}-F_{dis}^o}{T}=-\frac{F^{ex}_{dis}}{T},
\end{equation}
where $F^{ex}_{dis}$ is the {\em excess intrinsic free energy dissipation}.

Similar traditional and extended Second Law also hold beyond the level of free energy. Back to the evolution of entropy (\ref{eq_012}) and the nonnegativity of entropy production rate, the traditional Second Law here reads
$$\frac{d\tilde{S}^{open}}{dt}\geq -\frac{\tilde{h}_d^{open}}{T}.$$

One can further rewrite (\ref{eq_012}) as
\begin{equation}
   \frac{d\tilde{S}^{open}}{dt}=(e_p^{open}-\frac{F_{in}}{T})-\frac{\tilde{h}_d^{open}-F_{in}}{T}=\frac{F_{dis}^{net}}{T}-\frac{Q_{ex}}{T},\nonumber
\end{equation}
where the excess heat $Q_{ex}$ is the difference between $\tilde{h}_d^{open}$ and $F_{in}$ \cite{ep_fd_1}. The extended Second Law emerges \cite{ep_fd_1}
$$\frac{d\tilde{S}^{open}}{dt}\geq -\frac{Q_{ex}}{T}.$$
It is different from and stronger than the traditional Second Law, only for really driven system with $F_{in}=Q_{hk}>0$.

\section{Summary and discussion}

Is there anything beyond Boltzmann's notion about the Second Law of
Thermodynamics?  On this fundamental issue, J.L. Lebowitz \cite{lebowitz99RMP} and I. Prigogine \cite{thermo_np} seem to
disagree sharply: The former said ``no'' while the latter suggested
``yes''. Such a debate is focused on the evolution of entropy.
If we regard the entropy production $e_p$ as the total entropy increase of an isolated ``universe" and $h_d/T$ as the entropy change of the medium, then the fundamental equation in \cite{thermo_np} $dS/dt=e_p-h_d/T$ is nothing but a restatement of the entropy increase principle of any isolated system, long
realized by Helmholtz and Gibbs.  However, one could ask this question
differently: Is there anything beyond
Boltzmann's {\em microscopic} notion about the Second Law on a
{\em mesoscopic} scales?
One important notion of the NESS perspective \cite{QH}
is to build a self-consistent thermodynamics in terms solely of a mesoscopic,
Markov kinetics of open driven subsystems \cite{JQQ}, and to study
specially whether $h_d$ is really in the form of heat. This is the motivation
of the present article.

With the novel minimum work argument of idealized external regenerating
system, we show that the term $h_d$ is the minimum heat dissipation in an idealized NESS: It is also equal to the heat dissipation {\em plus} the entropy change of the environment due to the slowly changed external variables at QSS.
Furthermore, the total chemical and mechanical free energy input is balanced
by the $h_d$. In the light of this new perspective, the concept of efficiency
at NESS is well defined in terms of  energy {\em if} an outsider can
separately measure the chemical free energy input and mechanical
energy output.

A chemically driven system is referred to a physical or biological system with
a {\em sustained} source and sink with chemical potential difference.  The
concept, in fact, has a broader applicability to population dynamics than
traditional chemistry.  Fundamental physics
considers such a setup only approximate: In an absolutely sense,
the source and sink have to be slowly decay toward their own equilibrium (QSS).
Therefore, it is generally believed that if one includes these
relevant parts of ``environment'' into an enlarged system with detailed
balance, then the Prigogine's thesis would disappear.  However,
such a physical argument, which is absolutely valid, begs a resolution
on a mathematical level for the following paradox:
If one considers ($a$) NESS and QSS merely two
different perspectives of a same system; then it is surprising that they
have very different heat dissipation; if ($b$) one considers them as
different systems, then it is even more surprising their kinetics and
free energy accounting are identical.  We believe this is very much a
similar problem faced by Gibbs when he developed his different
ensemble theories for the {\em equation of equilibrium state}: Whether
one considers canonical and isobaric ensembles same or different, the
important issue is that they {\em both} give same macroscopic
thermodynamic relations; but they give different heat
capacity: $C_v$ and $C_p$.

The nonequilibrium thermodynamics of NESS and QSS actually
exhibit an {\em entropy-enthalpy compensation}, a phenomenon long observed in equilibrium statistical thermodynamics of complex biomacromolecular systems \cite{qian_hopfield,qian_jcp_98,qian_pre_01_eec,santillan_qian}:
Two very similar systems usually have a similar $\Delta\mu$,
but very different $\Delta h$ and $\Delta s$, and they compensate
$\Delta h-T\Delta s = \Delta\mu$.  This phenomenon was explained as
due to the fluctuations of different ensembles \cite{qian_hopfield,qian_jcp_98}.   Same phenomenon also arises in
stochastic kinetics of single macromolecules \cite{qian_pre_01_eec}
and in coarse graining stochastic thermodynamics \cite{santillan_qian}.
All these studies point to the highly malleable, yet powerful and efficient
concept of free energy and conditional free energy \cite{qian_preprint}.

Based
on established classical thermodynamics, we have obtained a
self-consistent picture for the nonequilibrium driven system. In fact, one can
see a distinction between Clausius' and Kelvin's
historical statements on the 2nd Law:
The former is about the spontaneity of a transient
process, i.e. the non-negativity of $f_d^{close}$, while
the latter is about a cyclic process
with non-negative $e_p^{open}$ in a NESS.

A new, mesoscopic position on the debate between the two different perspectives is now possible provided by recently developed purely mathematical
result on stochastic Markov process \cite{ep_fd_1,ep_fd_2,hq_decomp}.
In a nutshell, it has been shown, with some reasonable assumptions
and definitions, the entropy production rate $e_p$ consists of two
non-negative terms. Such a decomposition can be interpreted as
Boltzmann's thesis and Prigogine's thesis:
The former concerns with a system's spontaneous relaxation
to stationarity; and the latter concerns with a system that is sustained
at a nonequilibrium steady (stationary) state. This
suggests two origins of irreversibility at a mesoscopic level.

In the present article, we further refine this decomposition in a
stochastic transient state towards a NESS with an idealized external regenerating system:
The entropy production itself consists of the change of entropy
of the system, and total dissipated intrinsic free energy including that from the
minimally required amount from the regenerating system. The
house-keeping heat, on the other hand, consists of chemical and
mechanical free energies, as the net free energy input of the system.

All these are not only consistent with the traditional thermodynamics. They are also self-contained and solely dependent upon
only the internal kinetics of the NESS system, which can be directly measured
in statistical experiments. They also suggest new ingredients for a generalized nonequilibrium thermodynamics. As we have discussed, free energy is a powerful concept in systems with multiple scales, and  scientific theories indeed have a hierarchy structure, with new dynamics emerges at each specific collective level \cite{Anderson1972}.

We thank M. Esposito, C. Jarzynski, S. Sasa, U. Seifert, H.-Y. Wang for
helpful discussion. H. Ge is supported by the Foundation for the Author
of National Excellent Doctoral Dissertation of China (No. 201119).


\begin{thebibliography}{99}
\bibitem{footnote_1}
If ATP and ADP concentrations are sustained
at their equilibrium values, this setting is known as
{\em semi-grand ensemble} which is canonical for the motor
protein but grand-canonical for the ligands.
\end{thebibliography}
\end{document}